\documentclass[useAMS,usenatbib]{mn2e}
\usepackage{graphicx}
\usepackage{txfonts}
\usepackage{url}
\usepackage{color}
\usepackage{enumitem}
\usepackage{verbatim}
\usepackage{footnote}
\makesavenoteenv{tabular}
\usepackage[export]{adjustbox}
\makesavenoteenv{table}

\title[Constraining amC dust with 2MASS and \textit{Gaia}]{Optical properties of amorphous carbon dust around C-stars: new constraints from 2MASS and \textit{Gaia} observations}

\author[Nanni]{Ambra Nanni$^{1,2}$
  \\
$^1$ Dipartimento di Fisica e Astronomia Galileo Galilei,
 Universit\`a di Padova, Vicolo dell'Osservatorio 3, I-35122 Padova, Italy\\
$^2$ Aix Marseille Univ, CNRS, CNES, LAM, Marseille, France
}
\begin{document}

\date{Accepted 2018 November 06. Received 2018 November 04; in original form 2018 September 19}

\pagerange{\pageref{firstpage}--\pageref{lastpage}} \pubyear{2017}

\maketitle

\begin{abstract}\label{firstpage}
In this work the optical properties of amorphous carbon (amC) dust condensed around carbon(C)-stars are constrained by comparing the observations for the Large Magellanic Cloud C-stars from the Two Micron All Sky Survey (2MASS) and from the \textit{Gaia} data release 2 (DR2) with the synthetic photometry obtained by computing dust growth and radiative transfer in their circumstellar envelopes.
The set of optical constants of amC dust considered have been pre-selected according to their ability to reproduce the infrared colour-colour diagrams in the Small Magellanic Cloud. Only two combinations of the optical data set and grain size are able to reproduce the infrared photometry and the \textit{Gaia} observations simultaneously.
The analysis presented provides information about the properties of amC dust grains that might be characterized by a diamond-like structure, rather than a graphite-like one, at least around the most dust-enshrouded C-stars, or be composed of ``small'' grains of size less than 0.04~$\mu$m.
The selected data sets will be adopted to compute grids of spectra as a function of the stellar parameters that will be employed to estimate the dust return and mass-loss rates of C-stars by fitting their spectral energy distribution, and to study the resolved stellar populations of nearby objects.
\end{abstract}
\section{Introduction}\label{introduction}
In low-metallicity galaxies, such as the Magellanic Clouds (MCs), a large fraction of the thermally pulsing asymptotic giant branch (TP-AGB) stars are carbon(C)-rich. C-stars lose mass at high rates, returning metals and dust into the interstellar medium, and shaping the resolved stellar populations in the infrared bands.

Among the different dust species condensed in circumstellar envelopes (CSEs) of C-stars, amorphous carbon (amC) is normally the most abundant one. Since this dust species is very opaque to the stellar radiation, it largely affects the spectral energy distribution (SED) of C-stars.
In order to study such stars, it is therefore necessary to compute the dust radiative transfer through their CSEs.
In such a calculation the optical data set and grain size distribution of amC dust need to be chosen a priori. 
The optical properties of amC dust depend on the internal grain structure (more graphite-like or diamond-like), and several optical data sets are available in the literature \citep{Hanner88,Rouleau91,Zubko96,Jaeger98}. 
The choice of different combinations of the optical data set and grain size produce a sizeable effect in the emerging spectra and colours, especially for the most dust-enshrouded stars \citep{Nanni16}.

In \citet{Nanni16} the combinations of optical data sets and grain sizes that simultaneously reproduce the observed colour-colour diagrams (CCDs) of the Small Magellanic Cloud (SMC) in the infrared were selected. The synthetic photometry has been compared
with the observations from the Two Micron All Sky Survey \citep[2MASS;][]{skrutskie06} and from the \textit{Spitzer} Survey of the SMC \citep[S$^{3}$MC;][]{bolatto07}.
The calibration of the optical constants of amC dust is also valid for the Large Magellanic Cloud (LMC), since the trends in the infrared CCDs are similar to the ones in the SMC \citep{Boyer11}.

Recently, \citet{Lebzelter18} have developed a new technique to identify C-stars, by using a combination of Wesenheit functions in the near-infrared for the $J$ and $K_s$ bands of 2MASS, and in the optical, for the $G_{RP}$ and $G_{BP}$ bands of the \textit{Gaia} survey.
In the work here presented, the diagram introduced by \citet{Lebzelter18} is employed as a tool to further constrain the optical properties of amC dust condensed around C-stars, by comparing the observations with the synthetic photometry.
Constraining dust optical properties is relevant for interpreting different observations. Indeed, the spectra reprocessed by dust are adopted to simulate resolved stellar populations \citep{Marigo08, Marigo17}, and to estimate the mass-loss rates of individual stars, 
by fitting their SED \citep{Groenewegen09, Riebel12, Srinivasan16, Nanni17}.
\section{Method}\label{model}
\begin{table*}
\begin{center}
\caption{Input and output quantities of the dust growth calculations \citep{Nanni13,Nanni14,Nanni16, Nanni17}.}
\label{IO}
\begin{tabular}{l l}
\hline
Input  & Output \\
\hline
Mass-loss rate ($\dot{M}$) & Dust temperature at the inner boundary of the dust zone      \\
Current mass  &   Grain size, dust-to-gas ratio, dust mass-loss of each dust species, $i$ ($a_{\rm i}$, $\delta_{\rm i}$, $\dot{M}_{\rm i}=\dot{M}\delta_{\rm i}$)     \\
Effective temperature &  Total dust-to-gas ratio ($\delta=\sum_i \delta_i$) \\
Stellar luminosity &   Scattering, absorption and extinction efficiency for the dust mixture ($\bar{Q}_{\rm sca}$, $\bar{Q}_{\rm abs}$, $\bar{Q}_{\rm ext}$): $\bar{Q_{\rm k}}(\lambda)=\frac{\sum_i \dot{M}_{\rm i} Q_{\rm k, i}(a_{\rm i}, \lambda)}{\sum_i \dot{M}_{\rm i}}$\\
Elemental abundances at the photosphere & Outflow expansion velocity profile [$v_{\rm exp}(r)$] \\
Photospheric spectrum & Gas- and dust-density profiles [$\rho(r)$, $\rho_{\rm d}(r)$]: $\rho(r)=\frac{\dot{M}}{4\pi r^2 v_{\rm exp}(r)}$, $\rho_{\rm d}(r)=\rho(r)\delta$ \\
Seed particle abundance (equation \ref{seed_abundance})  & Condensation radius ($R_{\rm c}$) \\
Set of optical constants for each dust species & Optical depth as a function of the wavelength ($\tau_\lambda$): $\tau_\lambda=\frac{3\dot{M}}{4}\int_{R_{\rm c}}^{\infty} \sum_i \frac{Q_{\rm ext, i}(\lambda,a_i)}{a_i \rho_{\rm i}}\frac{\delta_i(r)}{r^2 v_{\rm exp}(r)}dr$ \\
\hline
\end{tabular}
\end{center}
\end{table*}
In the analysis presented, the Wesenheit function employed for the 2MASS filters are taken from \citet{Soszynski09}:
\begin{equation}
W_{K_s, J-K_s}=K_s-0.686\times(J-K_s).
\end{equation}
For the \textit{Gaia} data release 2 (DR2) bands the Wesenheit function has been obtained by \citet{Lebzelter18} from the interstellar extinction data provided by \citet{Schlegel98}:
\begin{equation}
W_{RP, BP-RP}=G_{RP}-1.3\times(G_{BP}-G_{RP}).
\end{equation}
The observed $K_s$ versus $W_{RP, BP-RP}-W_{K_s, J-K_s}$ index of C-stars in the LMC introduced by \citet{Lebzelter18} is compared with the synthetic photometry.

A dust-free stellar isochrone with metallicity of $Z=0.006$, suitable for the C-stars in the LMC, and $\log(\rm Age/yr)=8.8$ is computed with the code \textsc{trilegal} \citep{Girardi05}, using the public web interface \url{http://stev.oapd.inaf.it/cgi-bin/cmd}.
In \textsc{trilegal} the isochrones are computed from the stellar tracks calculated by the \textsc{parsec} code from the pre-main sequence up to the first thermal pulse \citep{Bressan12}, and by the \textsc{colibri} code during the TP-AGB \citep{Marigo13}. 

In the work presented here, dust growth and radiative transfer is computed for each of the models in the isochrone.
In the framework of this paper, grain growth is coupled with a spherical symmetric, stationary wind, as described in \citet{Nanni13, Nanni14}, which is a modified version of the formulation first introduced by \citet{FG06}. 
The original description by \citet{FG06} has also been applied to TP-AGB tracks in other works \citep{Ventura12,Ventura14, Ventura16,Dellagli15b, Dellagli15a}.

The input and output quantities of the dust growth code are listed in table \ref{IO}. The input parameters are the stellar quantities, the seed particle abundance,  $\epsilon_{\rm s, C}$, which  affects the final grain size \citep{Nanni17} and the dust optical constants.
Consistently with the previous works \citep{Nanni13, Nanni14, Nanni16, Nanni17}, the seed particle abundance is assumed to be proportional to the excess of carbon atoms with respect to the oxygen ones, 
\begin{equation}\label{seed_abundance}
\epsilon_{\rm s, C}\propto\epsilon_{\rm s}\times(C-O), 
\end{equation}
where $\epsilon_{\rm s}$ is an adjustable parameter.
Scaled-solar abundances for the elements are adopted in these calculations, with the exception of carbon, where it is determined by the number of carbon over the number of oxygen atoms in the stellar atmosphere ($C/O$).
For each model in the isochrone, the photospheric spectrum is interpolated in effective temperature and in $C/O$ between the ones of the \textsc{comarcs} grid \citep{Aringer09,Aringer16}. 
The metallicity of the photospheric spectra is about one third solar which is the value in the \textsc{comarcs} grid closest to the one adopted for the isochrone calculation.
Besides amC dust, grain growth is also computed for silicon carbide (SiC) and metallic iron. However, these two latter dust species are not expected to have a strong influence on the synthetic photometry considered in this work. Indeed, the amount of SiC and iron dust is relevant only for low values of the $C/O$ ratio ($C/O < 1.18$). For $C/O>1.18$ the dust mass fraction of SiC is up to $\approx$10\%, while iron dust is always $<1$\%.

The combinations of optical constants for the amC dust and $\epsilon_{\rm s}$ values in equation \ref{seed_abundance} are listed in table~\ref{opt} together with the size of amC dust grains obtained from the dust growth calculation applied to the isochrone. A complete description of the optical data sets is provided in \citet{Nanni16}. These combinations simultaneously reproduce the main CCDs for C-stars in the SMC in the infrared, and have been adopted to fit their SEDs and estimate their dust-production rates \citep{Nanni16,Nanni17}. 
The optical data sets for SiC and metallic iron are instead taken from \citet{Pegourie88} and \citet{Leksina67}, respectively. 
The dust absorption and scattering coefficients are computed assuming spherical grains, by means of the Mie code \textsc{bhmie} by \citet{Bohren83}.

Some of the quantities in table \ref{IO} are used as input for the radiative transfer calculation through the CSE performed with the code \textsc{dusty} \citep{Ivezic97}.
The input quantities for \textsc{dusty} are the effective temperature, the photospheric spectrum, the scattering and absorption efficiencies ($\bar{Q}_{\rm sca}$, $\bar{Q}_{\rm abs}$), the dust-density profile [$\rho_{\rm d}(r)$], the optical depth at a given wavelength ($\tau_\lambda$) and the dust temperature at the inner boundary of the shell.
All these quantities are consistently computed for the dust mixture and the dust-to-gas ratio calculated by means of the dust growth code.
This approach is different from the standard one in the literature, where the dust-to-gas ratio and dust chemical composition are fixed to a single value \citep{Groenewegen06, Groenewegen07, Groenewegen09, Gullieuszik12, Boyer12, Srinivasan16}.
Moreover, in the work here presented, the dust-density profile is not simply inverse-square as assumed in other works \citep{Groenewegen07,Srinivasan11}, but is computed taking into account the outflow acceleration (see table \ref{IO}). In particular, the expansion velocity profile of the outflow is not constant as instead assume in other works \citep{Groenewegen07,Groenewegen09, Boyer12,Riebel12}, but is derived from the calculation of dust growth coupled with the stationary wind (see also table \ref{IO}).
The emerging spectra reprocessed by dust obtained from the code \textsc{dusty} are normalized to their total luminosities \citep{Ivezic99}. The spectra are then convoluted with the 2MASS and \textit{Gaia} DR2 transmission curves \citep{Cohen03, Evans18} in order to obtain the synthetic photometry and the Wesenheit functions. The distance for the LMC is assumed to be $\sim 50$~kpc \citep{Ngeow08}.
\begin{table*}
\begin{center}
\caption{Combination of optical data sets and $\epsilon_{\rm s}$ values (see equation \ref{seed_abundance}) selected on the basis of \citet{Nanni16}. The corresponding grain sizes for amC dust ($a_{\rm amC}$) are the ones obtained by applying the dust growth calculation to the models in the stellar isochrone.}
\label{opt}
\begin{tabular}{c c c c c}
\hline
Optical data set  & $\rho_{\rm d, amC}$ [g cm$^{-3}$]& $\log(\epsilon_{\rm s})$ & $a_{\rm amC}$ [$\mu$m] &  Denomination \\
\hline
\citet{Rouleau91} & 1.85 & $-12$ & $0.004$--$0.08$  &R12 \\
\citet{Rouleau91}  & 1.85 & $-13$ & $0.004$--$0.17$ & R13 \\
\citet{Jaeger98} (T=400 $^\circ$C)  & 1.435 &$-12$ & $0.001$--$0.08$ & J400 \\
\citet{Jaeger98} (T=1000 $^\circ$C) & 1.988 &$-12$   & $0.004$--$0.07$ &J1000 \\
\citet{Hanner88} &  1.85 & $-11$ & $0.004$--$0.04$  &H11 \\
\citet{Zubko96} (ACAR sample) & 1.87 &$-12$ & $0.004$--$0.08$ &Z12\\
\hline
\end{tabular}
\end{center}
\end{table*}
\section{Selected sample of stars} 
The $J$ and $K_s$ photometry is taken from the list of TP-AGB candidates provided by \citet{Riebel12}.
In this catalogue, stars are classified as C- or oxygen-rich according to their location on the $K_s$ versus $J- K_s$ colour--magnitude diagram \citep{cioni06a}. The most extreme (X-), dust-enshrouded stars are classified by employing the observations in the mid-infrared bands from the \textit{Spitzer} surveys of the LMC \citep[SAGE;][]{blum06}. 
The majority of the X-stars are probably carbon-rich \citep{vanLoon97, vanLoon06, vanLoon08, Matsuura09}.

The list of TP-AGB candidates by \citet{Riebel12} is matched within 1$^{\prime\prime}$ with the catalogues of spectroscopically classified sources by \citet{Jones17} and \citet{Groenewegen18}.
Matches are found for 246 sources (151 C-stars) in the catalogue by \citet{Jones17} and for 227 sources (159 C-stars) in \citet{Groenewegen18}.
The stars spectroscopically identified as C-stars by \citet{Groenewegen18} or, in case this is not available, by \citet{Jones17}, are included in the sample together with the ones classified as C- or X-stars by \citet{Riebel12} that lack a spectroscopic identification. The stars spectroscopically classified as oxygen-rich are excluded from the sample.
In the catalogue by \citet{Riebel12} 6709 sources are photometrically classified as C-stars and 1352 as X-stars, for a total number of 8061 stars. 
Among this sample, 4 and 12 sources identified as C- and X-stars respectively by \citet{Riebel12} are spectroscopically classified as oxygen-rich according to the criteria employed. 
On the other hand, 5 sources classified as oxygen-rich by \citet{Riebel12} are instead C-stars on the basis of the spectroscopic classification.
However, these latter stars do not have valid photometry in the $J$ and $K_s$ bands in both the catalogues by \citet{Riebel12} and \citet{Jones17}.
Furthermore, 27 sources spectroscopically classified as carbon-rich stars are not included in the catalogue by \citet{Riebel12}.
However, these sources do not have valid 2MASS photometry in the catalogue by \citet{Jones17} and consequently they are not studied.

The sample selected is composed by 8045 stars, of which 6705 are C-stars and 1340 are X-stars.
For 11 X-stars the 2MASS photometry is not listed in the catalogue by \citet{Riebel12} and the one from \citet{Jones17} is selected.

The sample obtained is cross-matched within 1$^{\prime\prime}$ with the \textit{Gaia} photometry from DR2 \citep{Gaia16,Gaia18,Riello18,Evans18} 
by employing the tool available at \url{http://gea.esac.esa.int/archive/}.
The final sample of stars with valid photometry in both in the 2MASS and \textit{Gaia} bands is composed by 6677 C-stars and 894 X-stars which represent $99.5$\% and $66.7$\% of the two classes of stars selected from \citet{Riebel12} as previously described.

\section{Results}
\begin{figure*}
\includegraphics[scale=0.9]{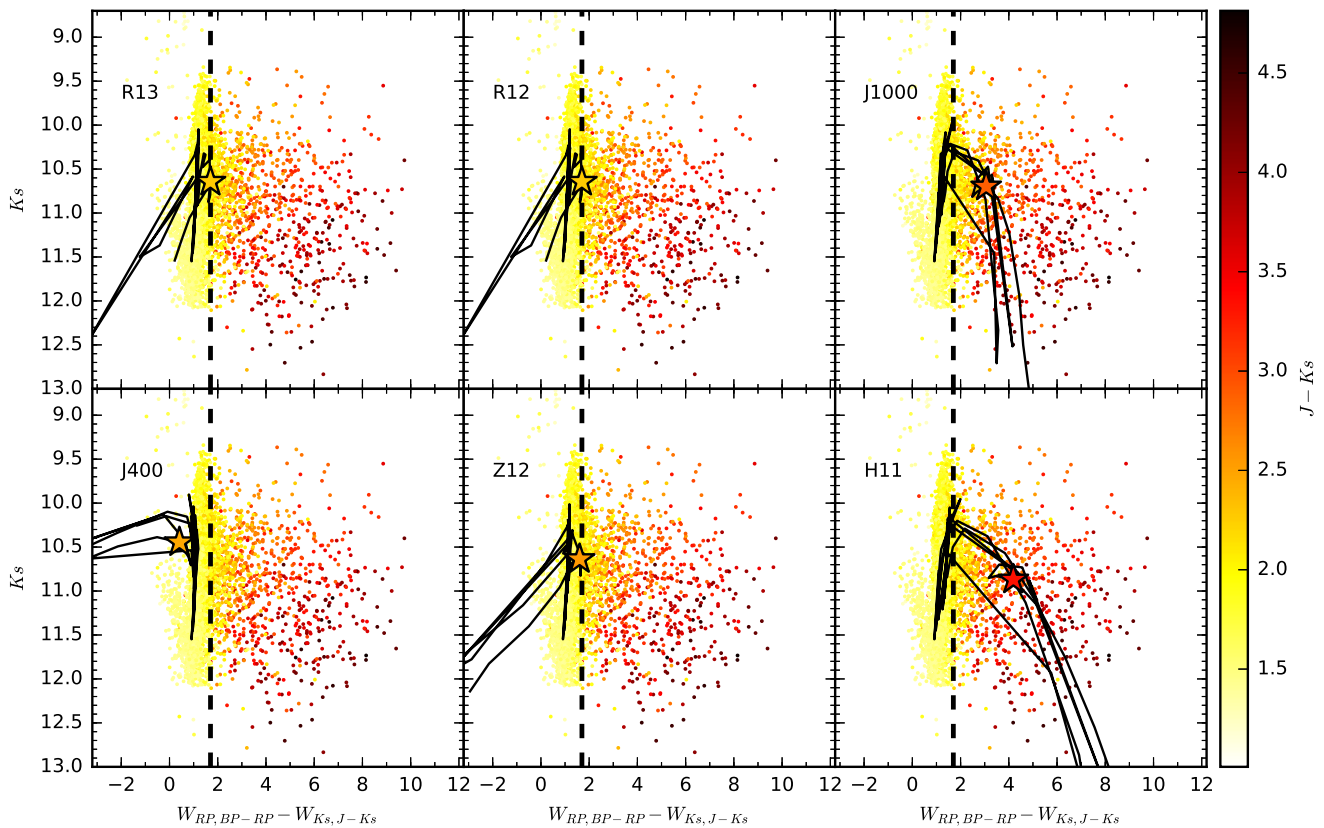}
\caption{$K_s$ from 2MASS as a function of the $W_{RP, BP-RP}-W_{K_s, J-K_s}$ index (small dots). The synthetic photometry computed for a selected isochrone by employing the different optical constants listed in table~\ref{opt} is overlaid onto the observations (solid-black line). The same model in the isochrone is indicated with a star-like symbol in the different panels. The observations and the selected model are colour-coded according to their $J-K_s$ value. The vertical dashed line indicates the separation between C- and X-stars according to \citet{Lebzelter18}.}
\label{fig:all_index}
\end{figure*}

In figure \ref{fig:all_index} the $K_s$ against the $W_{RP, BP-RP}-W_{K_s, J-K_s}$ index is shown for the C-stars in the LMC. The synthetic photometry in the isochrone computed for the six optical data sets in table~\ref{opt} is overlaid onto the observations. 
A selected model with mass-loss rate of $\sim4.6\times10^{-6}$ M$_{\odot}$yr$^{-1}$ and $C/O\sim 4.9$ is also highlighted with a star-like symbol.
Both the observed stars and simulated photometry are colour-coded according to their $J-K_s$ value.

For the observed C-stars, the $W_{RP, BP-RP}-W_{K_s, J-K_s}$ index becomes redder for increasing values of the $J-K_s$ colour.
On the other hand, the synthetic models show two distinct behaviours: either they move to the left or to the right part of the plot, as they become more dust-enshrouded.
Only the isochrones derived from the H11 and J1000 optical data sets move to the right, in agreement with the observed trends.
The $J-K_s$ colour of the model highlighted in the plot is between $2.3$ and $3.3$ mag, depending on the selected optical data set.
For the H11 and the J1000 such a model is located around $W_{RP, BP-RP}-W_{K_s, J-K_s}\sim3$ mag and $W_{RP, BP-RP}-W_{K_s, J-K_s}\sim 4$ mag, respectively. For the H11 and the J1000 optical data sets, the $J-K_s$ colour is in fairly good agreement with the ones of the observed stars that are placed in the same area of the diagram.

For R13, R12 and Z12 the simulated star is located around $W_{RP, BP-RP}-W_{K_s, J-K_s}\sim 2$ mag, while for J400, $W_{RP, BP-RP}-W_{K_s, J-K_s}\sim 1$ mag. The model computed for the Z12 and for the J400 data sets is located in a region of the plot corresponding to mildly dust-enshrouded stars, and its $J-K_s$ colour is not in good agreement (too red) with the ones observed in the same area of the diagram.
On the other hand, the $J-K_s$ colours computed with the R12 and the R13 data sets, are only slightly redder than the ones observed in the same area of the diagram, but the isochrone never becomes red in $W_{RP, BP-RP}-W_{K_s, J-K_s}$, as it is instead observed.

\begin{figure*}
\includegraphics[scale=0.9]{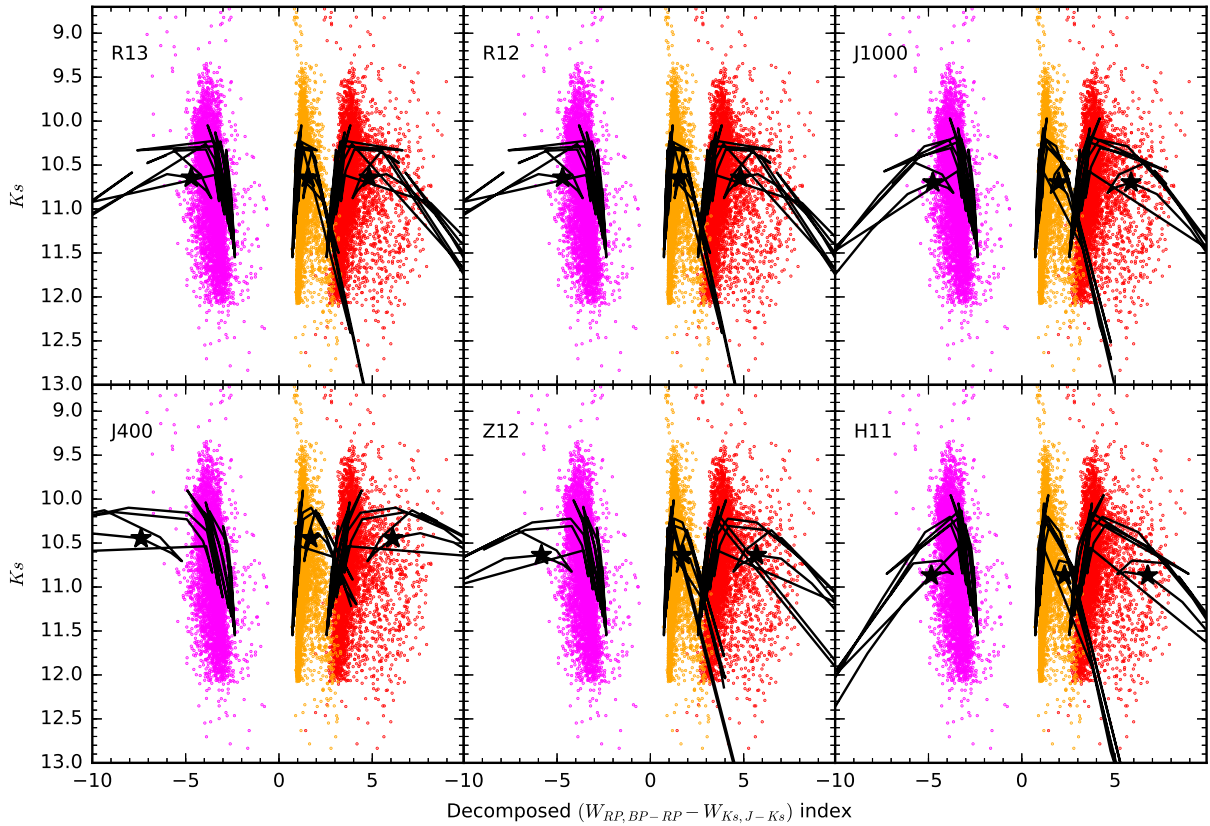}
\caption{$K_s$ from 2MASS as a function of the $W_{RP, BP-RP}-W_{K_s, J-K_s}$ index (small dots) decomposed in three components: $-1.3\times(G_{BP}-G_{ RP})$, left sequence in magenta, $+0.686\times(J-K_s)$, central sequence in orange and $G_{RP}-K_s$, right sequence in red. The same isochrone and the same model of figure \ref{fig:all_index} computed by employing the optical data sets in table \ref{opt} are plotted with a solid-black line and a star-like symbol, respectively.}
\label{index_comp}
\end{figure*}

In order to understand why different combinations of optical constants and grain sizes behave differently, the $W_{RP, BP-RP}-W_{K_s, J-K_s}$ index has been decomposed in three components: $-1.3\times(G_{BP}-G_{RP})$, $+0.686\times(J-K_s)$ and $G_{RP}-K_s$, as done in \citet{Lebzelter18}.
The results are shown in figure \ref{index_comp} for all the selected optical constants. The same dust-enshrouded star as in figure \ref{fig:all_index} is shown in the plot. 
The model highlighted lies in the observed sequences given by $K_s$ versus $+0.686\times(J-K_s)$ and by $K_s$ versus $G_{RP}-K_s$ for all the optical constants.
The values of $G_{RP}-K_s$ attained by the H11, J1000, J400 and Z12 data sets are redder than the ones obtained with the R12 and the R13 sets by about 1 mag.
For the selected point along the isochrone, the $K_s$ versus $-1.3\times(G_{BP}-G_{RP})$ is in agreement with the observations for all the optical constants but for the J400 and Z12 sets for which the theoretical value is bluer than the ones observed by $\sim 2.5$ mag and $\sim 1$ mag, respectively.
By adding up the three contributions shown in figure \ref{index_comp} it is therefore possible to understand why the same model computed by employing different optical constants is located in different part of the $K_s$ versus $W_{RP, BP-RP}-W_{K_s, J-K_s}$ diagram in figure \ref{fig:all_index}.

\begin{figure}
\includegraphics[width=0.48\textwidth]{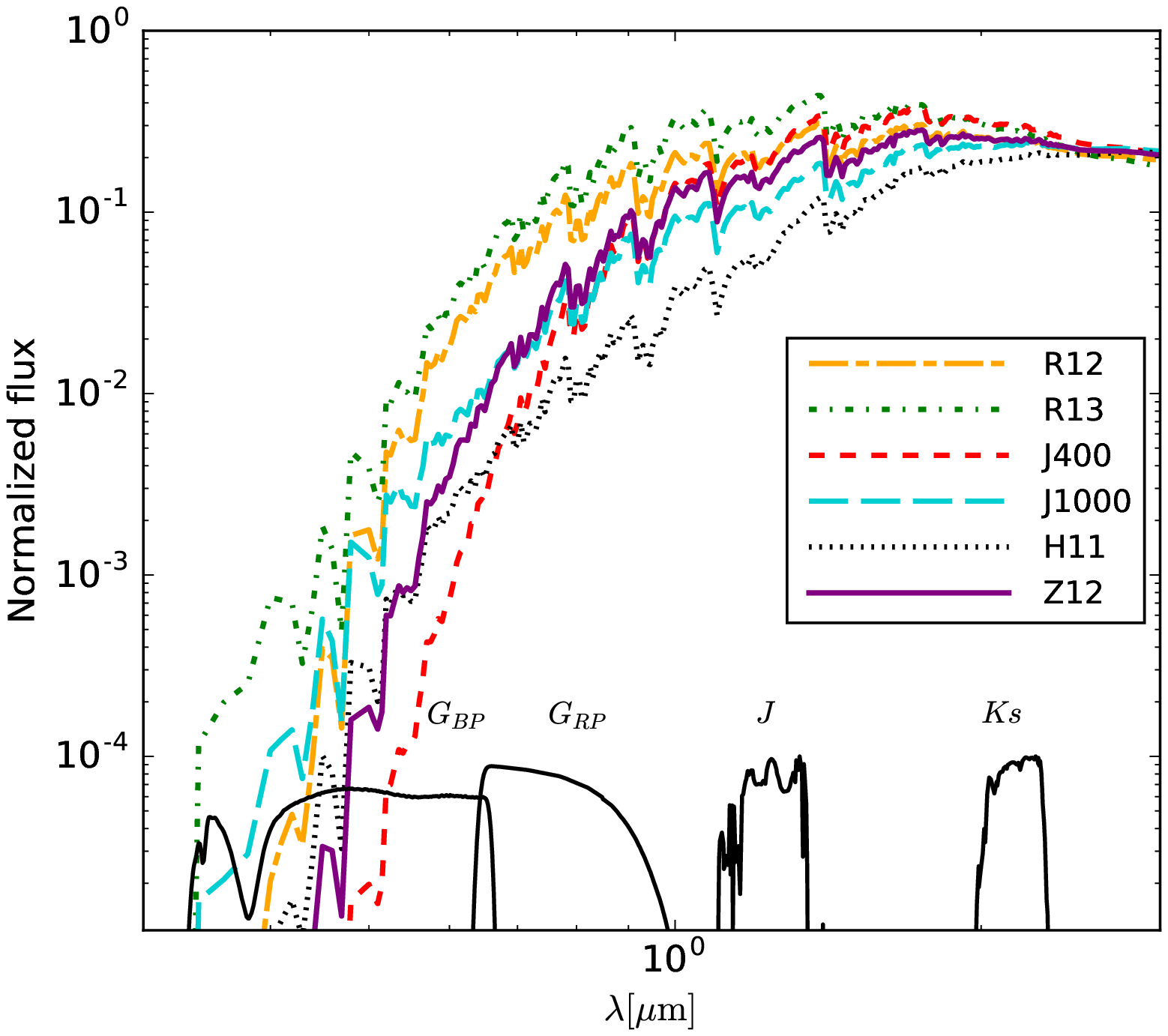}
\includegraphics[width=0.48\textwidth]{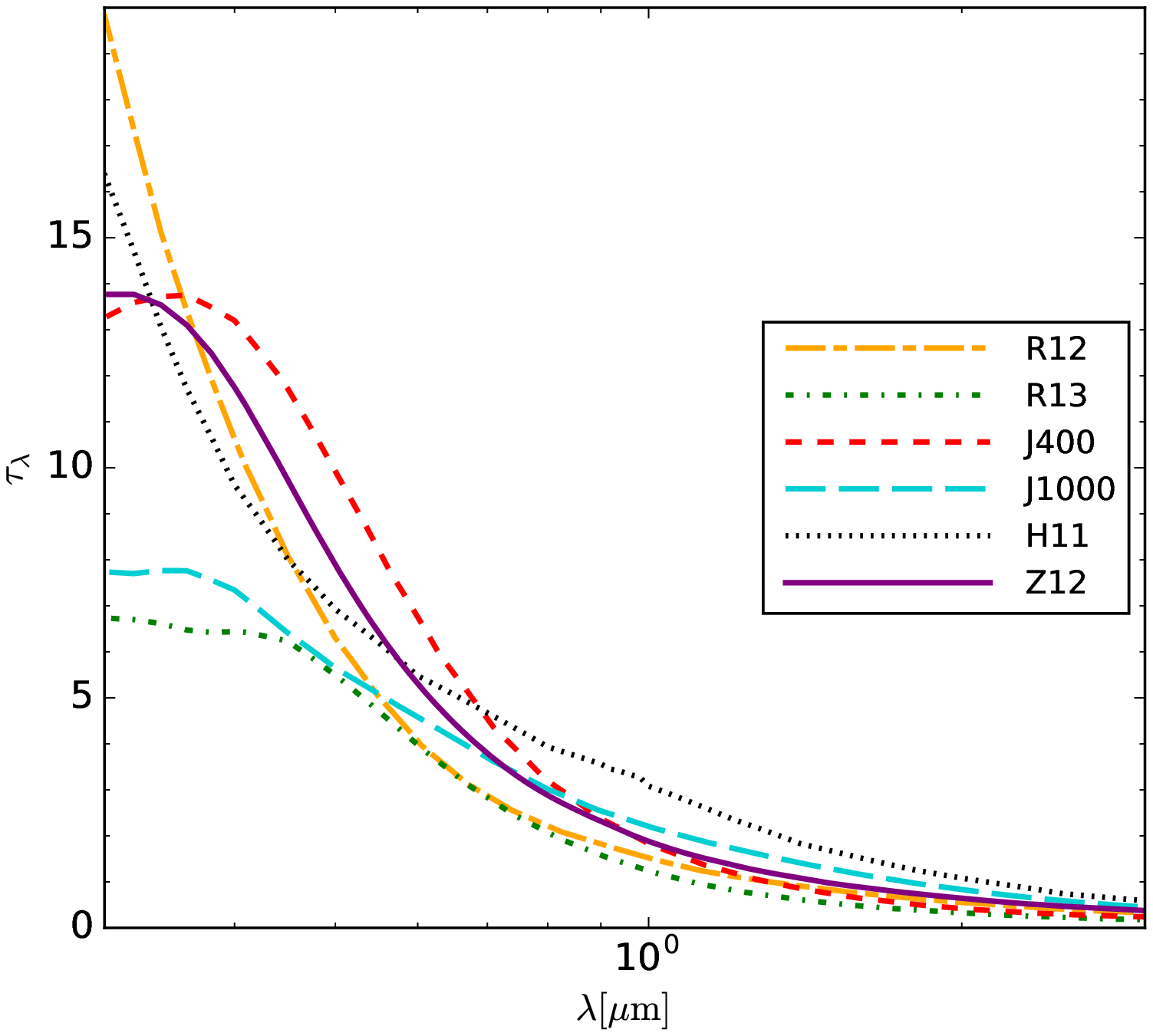}
\caption{Upper panel: spectra of the same model in the isochrone indicated with a star-like symbol in figures \ref{fig:all_index} and \ref{index_comp} computed by employing the  different optical data sets. Different colours and styles in the legend indicate the set of optical constants adopted for each calculation. The transmission curves divided by one thousand for the \textit{Gaia} DR2 and 2MASS filters used in this work are also shown. Lower panel: optical depth as a function of the wavelength (the same colour code is adopted).}
\label{Gaia_spectra}
\end{figure}

The behaviour of the different components of the Wesenheit function for the model selected in the isochrone reflects the differences in the spectra computed with different optical data sets, as shown in figure \ref{Gaia_spectra}.
The emerging spectra obtained for the model highlighted in figures \ref{fig:all_index} and \ref{index_comp} by employing the different optical data sets are plotted in the upper panel of figure \ref{Gaia_spectra}. The response curves for the 2MASS and \textit{Gaia} filters used in this work are also shown.
In the lower panel of the same figure the corresponding $\tau_\lambda$ are plotted.
The shape of $\tau_\lambda$ is determined by the absorption and scattering efficiencies and by the grain size, while the dust-density profile sets its absolute value (see the corresponding equation in table \ref{IO}). The quantity $\tau_\lambda$ mainly determines the emerging spectra reprocessed by dust.
As an example, the spectra computed for the J400 and for the H11 data sets, for which the model in the isochrone shows opposite trends in figure \ref{fig:all_index}, are discussed.
The ratio between the fluxes in the $G_{BP}$ and $G_{RP}$ bands is lower for the J400 data set, and therefore its $-1.3\times(G_{BP}-G_{RP})$ colour is bluer than the one obtained with the H11. 
On the other hand, the flux in the $K_s$ band is approximately the same for the two data sets, but it is larger in the $J$ and in the $G_{RP}$ bands for the J400. As a consequence, the J400 model is slightly bluer than the H11 one in the $+0.686\times(J-K_s)$ and $G_{RP}-K_s$ colours in figure \ref{index_comp}. 
With similar arguments is possible to explain the synthetic colours of the model shown in figure \ref{index_comp} for all the optical constants employed in the calculations presented here.

By analysing the two sets of optical constants that satisfactorily reproduce the observations, the J1000 and the H11, it is possible to derive relevant information about the amC dust properties.
\begin{itemize}
\item \textit{J1000}. \citet{Jaeger98} have measured the optical constants of different amC substrates, obtained by pyrolyzing cellulose material at different temperatures. The J1000 and the J400 optical data sets have been obtained from the measurements of the optical constants of material pyrolized at 1000$^\circ$C and at 400$^\circ$C, respectively.
The substrates obtained are characterized by different internal structures: more diamond-like or graphite-like, according to their pyrolization temperatures. In particular, the material becomes more diamond-like for increasing pyrolization temperatures. 

The J1000 data set with grain sizes smaller than $0.07$~$\mu$m is in agreement with the infrared observations, and produces good results also in the $K_s$ versus $W_{RP, BP-RP}-W_{K_s, J-K_s}$ plane.
This outcome might indicate that amC dust grains formed in CSEs of dust-enshrouded C-stars are characterized predominantly by a diamond-like structure. This result is in agreement with the finding by \citet{Nanni16}, in which a better agreement with the infrared observations of the most dust-enshrouded C-stars was obtained by employing the J1000 data set rather than the J400 one.
\item \textit{H11}. The H11 optical data set from \citet{Hanner88} is based on laboratory measurements of the optical constants of glassy amC by \citet{Edoh83}.
For the H11 data set the grain sizes,  $\lesssim0.04$~$\mu$m, are the smallest ones obtained from the analysis by \citet{Nanni16}.
This result might indicate that the typical grain size of amC dust grains around C-stars can be at least $2.5$ times smaller than the value usually adopted in radiative transfer calculations, which is $0.1$~$\mu$m \citep{Groenewegen09,Srinivasan11}. 
\end{itemize}
Some degeneracy in these results related to the grain size and/or internal structure of amC dust might be possible when different grain configurations are considered, i.e. hollow spheres or core-mantle grains. However, this analysis is beyond the scope of this work and can be the subject of future investigations.

\section{Conclusions}
The analysis presented in this work allows to better constrain on the optical properties of amC dust, by employing the observations in the optical bands from \textit{Gaia} DR2.
The optical data sets considered in this work are the ones able to reproduce the observations in the infrared bands \citep{Nanni16}. 
In the current investigation, the 2MASS and the \textit{Gaia} photometry are combined to provide in a specific diagram, $K_s$ versus $W_{ RP, BP-RP}-W_{ K_s, J-K_s}$, introduced by \citet{Lebzelter18}, that is compared with the synthetic photometry calculated for a specific isochrone.
The theoretical isochrones show two different behaviours in $K_s$ versus $W_{ RP, BP-RP}-W_{ K_s, J-K_s}$ plane: either they move to the right or to the left part of the diagram when they become more dust-enshrouded.
Since the observed C-stars become redder in the $W_{ RP, BP-RP}-W_{ K_s, J-K_s}$ index as they become more dust-enshrouded, only the models that move to the right part of the diagram are in agreement the observations.

Two of the data sets among those that well fit the CCDs in the infrared are also able to reproduce the observed trends in the $K_s$ versus $W_{RP, BP-RP}-W_{K_s, J-K_s}$ diagram: 
\begin{itemize}
\item The J1000 optical data set by \citet{Jaeger98} with grains of size $\lesssim0.07$~$\mu$m. This result suggests that amC dust condensed around dust-enshrouded C-stars might be characterized by a diamond-like structure, in agreement with the analysis by \citet{Nanni16}.
\item The H11 data set by \citet{Hanner88} with grains of size $\lesssim0.04$~$\mu$m.
The maximum grain size achieved by these models is at least $2.5$ times smaller than the value usually assumed in the literature, which is $0.1$~$\mu$m \citep{Groenewegen09,Srinivasan11}. 
\end{itemize}
The two aforementioned optical data sets will be employed to calculate large grids of spectra for C-stars as a function of the stellar parameters that will be adopted to fit the SED of the C-stars in the MCs. This will allow to estimate their dust-production rate as well as their mass-loss rates, as already done in \citet{Nanni17}. Restricting the analysis to two optical data sets will allow to provide a better constraint on the total dust-production rate of C-stars in the MCs \citep{Nanni17}.

Since the set of optical constants selected affects the wind dynamics, the calibration presented here will allow to interpret the wind speeds derived for C-stars from CO line measurements both in the Milky Way \citep{Ramstedt14, Danilovich15}, after testing the method at solar metallicity, and in the LMC, for which the wind speed measurements have been performed with the Atacama Large Millimeter Array \citep{Groenewegen16}. 
The grids of spectra can also be employed to simulate resolved stellar populations in order to interpret the most recent observations from \textit{Gaia} DR2.
\subsection*{Acknowledgements}
This work was supported by the ERC Consolidator Grant funding scheme
({\em project STARKEY}, G.A. n.~615604) and by the Centre National d'{\'E}tudes Spatiales (CNES).

I thank all the members of the STARKEY group and Professor Alessandro Bressan for the fruitful discussions.
I thank the anonymous referee for carefully reading the manuscript and for the useful suggestions that improved this work.

This work has made use of data from the European Space Agency (ESA) mission
{\it Gaia} (\url{https://www.cosmos.esa.int/gaia}), processed by the {\it Gaia}
Data Processing and Analysis Consortium (DPAC,
\url{https://www.cosmos.esa.int/web/gaia/dpac/consortium}). Funding for the DPAC
has been provided by national institutions, in particular the institutions
participating in the {\it Gaia} Multilateral Agreement.

\bibliographystyle{mn2e_new}
\bibliography{nanni}

\label{lastpage}
\end{document}